\documentstyle[preprint,aps]{revtex}

\begin{document}
\draft
\title{End states, ladder compounds, and domain wall fermions}
\author{Michael Creutz
}
\address{
Physics Department, Brookhaven National Laboratory,\\
Upton, NY 11973
}
\date{\today}
\maketitle
\begin{abstract}

A magnetic field applied to a cross linked ladder compound can
generate isolated electronic states bound to the ends of the chain.
After exploring the interference phenomena responsible, I discuss a
connection to the domain wall approach to chiral fermions in lattice
gauge theory.  The robust nature of the states under small variations
of the bond strengths is tied to chiral symmetry and the
multiplicative renormalization of fermion masses.

\end{abstract}


\narrowtext

%\section{Introduction}

Isolated electronic states can appear when a magnetic field is applied
to a ladder system with diagonal cross linking.  The basic structure
is shown in Fig.~\ref{fig:icetray2}.  A single electron inserted in
such a system will hop around and, with dissipation, settle into a
ground state with wave function non-vanishing on all sites.  When a
magnetic field is present, phases appear which can cause interesting
interference effects.  In special cases this results in unusual states
that do not diffuse through the system, but instead are bound to the
ends of the chain.  When the field has a strength of half a flux unit
per plaquette, symmetries exactly determine the energy of these
special states.

These end states are a particular manifestation of the surface states
discussed by Shockley \cite{shockley}.  They are presumably also a one
dimensional remnant of the edge states discussed in
Ref.~\cite{edgestates}.  My entire treatment is in terms of free
fermions hopping through the fixed lattice; I include no four fermion
interactions.

On the promotion of the system to a five space-time dimensional model
with appropriate spin factors, these states become precisely the
chiral modes of the Kaplan-Furman-Shamir \cite{dwf} domain-wall
fermions currently being used \cite{dwfrecent} in lattice gauge
simulations.  My goal here is a qualitative picture for the chiral
nature of these surface modes.

%\section{The ladder model}

I start with a spin-less electron on the basic structure of
Fig. \ref{fig:icetray2}.  All horizontal bonds are of equal strength,
given by hopping parameter $K$.  For vertical bonds, the coupling is
$M$ and for the diagonal bonds $Kr$.  I assume these parameters are
positive.  The notation is motivated by the later connection with
lattice gauge models.  Thus my starting Hamiltonian is
\begin{eqnarray}
H=-\sum_n \bigg(&&
K(a_n^\dagger a_{n+1}+a_{n+1}^\dagger a_n
+b_n^\dagger b_{n+1}+b_{n+1}^\dagger b_n )\cr
&&+Kr(a_n^\dagger b_{n+1}+b_{n+1}^\dagger a_n
+b_n^\dagger a_{n+1}+a_{n+1}^\dagger b_n) \cr
&&+M(a_n^\dagger b_{n}+ b_n^\dagger a_{n})\bigg)
\end{eqnarray}
where $a_n$ and $b_n$ are fermionic annihilation operators on the
$n$'th site of the upper and lower chains, respectively.  A general
one particle state is
\begin{equation}
|\chi\rangle = \sum_{n=1}^L \pmatrix{a_n^\dagger & b_n^\dagger\cr}
\pmatrix{\psi^a_n \cr \psi^b_n} |0\rangle
\end{equation}
Here $|0\rangle$ represents the bare vacuum with no fermions present.

For an infinite chain this model is easily solved via Fourier
transform 
\widetext
\begin{equation}
H=-
\int_0^{2\pi} {dq\over 2\pi} 
\pmatrix { \tilde a_q^\dagger & \tilde b_q^\dagger \cr}
\pmatrix { 2K\cos(q) & M + 2Kr\cos(q) \cr
M+2Kr\cos(q) & 2K\cos(q) \cr}
\pmatrix{ \tilde a_q \cr \tilde b_q \cr}
\label{eq:mom1}
\end{equation}
\narrowtext 
\noindent where $ \tilde a_q = \sum_n a_n e^{iqx}$.  Thus the energy
eigenvalues are
\begin{equation}
E_\pm(q)=-2K\cos(q) \pm \left(M+2Kr \cos(q)\right)
\end{equation}
Generically, two bands correspond to even and odd states under
inversion of the system around the horizontal axis.  Depending on
parameters, these bands may or may not overlap.  Here I am interested
in the small $M/K$ case, where they do overlap.  These momentum space
solutions carry over to a finite periodic system of length $L$, except
with discrete values for the momentum, $q=2\pi n/L$, $n$ an integer.

Now apply a perpendicular magnetic field.  This induces phases when a
particle hops around closed loops.  The details are gauge dependent; I
adopt the convention of placing these factors on the upper and lower
horizontal bonds, generalizing my Hamiltonian to \widetext
\begin{eqnarray}
H=-\sum_n \bigg(&&
K(e^{i\theta}a_n^\dagger a_{n+1}+e^{-i\theta}a_{n+1}^\dagger a_n+e^{-i\theta}b_n^\dagger b_{n+1}+e^{i\theta}b_{n+1}^\dagger b_n )\cr
&&+Kr(a_n^\dagger b_{n+1}+b_{n+1}^\dagger a_n 
+b_n^\dagger a_{n+1}+a_{n+1}^\dagger b_n) \cr
&&+M(a_n^\dagger b_{n}+ b_n^\dagger a_{n})\bigg)
\end{eqnarray}
In natural units, the magnetic flux is $\theta/\pi$ through each
plaquette.  Eq.~\ref{eq:mom1} now becomes
\begin{equation}
H=-
\int_0^{2\pi} {dq\over 2\pi} 
\pmatrix { \tilde a_q^\dagger & \tilde b_q^\dagger \cr}
\pmatrix { 2K\cos(q-\theta) & M + 2Kr\cos(q) \cr
M+2Kr\cos(q) & 2K\cos(q+\theta) \cr}
\pmatrix{ \tilde a_q \cr \tilde b_q \cr}
\label{eq:mom2}
\end{equation}
\noindent
and the energy eigenvalues are
\begin{equation}
%\begin{eqnarray}
E_\pm(q) 
%&&
= -2 K\cos(q) \cos(\theta) 
%\cr &&
\pm 
\sqrt{ 4K^2\sin^2(q) \sin^2(\theta) + (M+2Kr\cos(q))^2}
\label{eq:spectrum}
%\end{eqnarray}
\end{equation}
\narrowtext 
For sufficient field, the two bands can separate, leaving
a gap in the spectrum.  For small $M/2Kr$, this opening can produce
the situation discussed in Ref.~\cite{shockley}.  For small $r$ the
lower band has its maximum at $q=\pi$, the upper band has its minimum
at $q=0$ and the band separation occurs at $\theta = \cos^{-1}(r)$.
For large $r$, the separation of the bands occurs at momentum
$q=\pi-\cos^{-1}(M/2Kr)$, as soon as the field is present.

A particularly interesting limit occurs for half a flux unit per
plaquette, {\it i.e.} $\theta=\pi/2$.  These relative phases are
sketched in Fig.~\ref{fig:icetray2}.  The cancellations are most
dramatic when $M=0$ and $r=1$, i.e. no vertical bonds and all others
equal in magnitude.  Then all paths from a given site to any location
two or more sites away cancel, as illustrated in
Fig.~\ref{fig:cancel}.  Electrons cannot diffuse throughout the
system.  Eq.~\ref{eq:spectrum} reduces to all states having either
energy $2K$ or $-2K$.  With constant energies, group velocities vanish
and nothing moves.  For explicit solutions, consider a given site $n$
and the (unnormalized) wave function with $\psi_n=\pmatrix {i\cr
1\cr}$, $\psi_{n+1}=\pm\pmatrix {1\cr i\cr}$ and all other $\psi_j=0$.
These soliton like states have energy $\mp 2K$, respectively and are
locked onto the plaquette between site $n$ and $n+1$.  As there are
two solutions for every plaquette and there twice as many atoms as
plaquettes, this is a complete set of states for either the infinite
or the periodic case.

%\section{The end states}

Now suppose we have not a periodic but a finite open system.  Let the
sites (each with two atoms) run over the range $1\le n \le L$.  Open
boundaries means no direct connection from site $L$ back to site $1$.
For the moment, I stay with the above special case with
$\theta=\pi/2$, $M=0$ and $r=1$.  Now there is one more site than the
number of plaquettes, and therefore two more states than the above
solitons.  One of these states has
\begin{equation}\matrix{
&\psi_1=\pmatrix{1 \cr i \cr}\cr
&\psi_n=0\qquad n\ne 1 \cr
}
\end{equation}
The electron cannot hop anywhere because all such attempts interfere
completely.  It is locked onto the end of the chain.  This state is
time independent; it has exactly zero energy.  A symmetric state
binds to the other end of the system, with
\begin{equation}\matrix{
&\psi_L=\pmatrix{i \cr 1 \cr}\cr
&\psi_n=0\qquad n\ne L \cr
}
\end{equation}
These two  states are related by  rotating by $\pi$ about  the axis of
the magnetic field.

Now move away from the special case with $M=0$ and $r=1$.  Then the
upper and lower bands broaden and the above solitonic states mix and
move.  By continuity, however, the end states cannot suddenly
disappear.  They are robust under small variations of the parameters.
Keeping $\theta=\pi/2$ and considering the long $L$ limit, symmetries
force these modes to remain at zero energy.  Rotating the system by
$\pi$ about the the magnetic field interchanges them; thus they have
equal energy.  However, redefining all the $b$ operators to include a
minus sign, i.e.  $b_n\rightarrow -b_n$, and then swapping the ends of
the system with $n\rightarrow L-n+1$ takes the Hamiltonian into its
negative.  This requires the energies to be of opposite sign.  In
particle physics language, the energies are protected from
renormalization.  This is reminiscent of the role of chiral symmetry
in continuum field theories, and is the basis of the domain wall
analogy discussed below.

This restriction on the energies weakens when the magnetic field
strength varies.  The left-right symmetry keeps the states degenerate,
but no longer necessarily at zero energy.  As the field is reduced,
generically the states drift until pinched between the bands of the
full system.  For the remainder of this discussion I keep
$\theta=\pi/2$, and the end states remain at zero energy.  While
taking $(M,r)=(0,1)$ seems to avoid the requirement of a large $L$
limit, this is not robust.  Indeed, to avoid doublers, these
parameters become functions of physical momenta when I later extend
the model to higher dimensions.

To be more explicit, consider small $M$, keeping $\theta=\pi/2$ and
$r=1$.  Then the solution remains quite simple in the large $L$ limit,
{\it i.e.} for a semi-infinite chain we have the (unnormalized) wave
function
\begin{equation}
\psi_n=(-M/2)^n\pmatrix{1 \cr i \cr},
\label{exponential}
\end{equation}
sketched in Fig.~\ref{fig:icetray3} for $M=1$.  The modes extend into
the chain with an exponential falloff.  Note that when $M$ gets too
large, i.e. at $M=2$, the modes are lost.  At this point the wave
function is no longer damped, and the modes are absorbed into the
continuous bands.  This is also the point at which the band gap
vanishes.  Finally, for finite $L$ these modes feel the opposite ends
of the system.  Generically a small mixing, exponentially suppressed
with the length $L$, drives them from exactly zero energy.

Varying $r$ preserves the modes as well, although the
resulting wave function becomes a bit more complicated.
Although the notation is different, the
following discussion is essentially equivalent to arguments
in Ref.~\cite{mcih}.  The desired solution is the superposition of two
eigenstates under translation along the lattice.  The boundary
condition for the wave function to vanish at site 0
can be guaranteed with the form
\begin{equation}
\psi_n={\lambda_1^n-\lambda_2^n \over \lambda_1-\lambda_2} \psi_1
\end{equation}
Thus we need the peculiar situation of two different translation
eigenvalues with the same eigenvector.  At zero energy the basic
requirement for a translation eigenvalue is
\begin{equation}
\pmatrix{ -i(1-\lambda^2) & r(1+\lambda^2)+(M/K)\lambda \cr
          r(1+\lambda^2)+(M/K)\lambda & i (1-\lambda^2) \cr}
\psi=0
\end{equation}
Then $\psi=\pmatrix{1 \cr i \cr}$ is an eigenvector for both
solutions of the quadratic equation
\begin{equation}
r(1+\lambda^2)+(M/K)\lambda = 1-\lambda^2 
\end{equation}
These two solutions are
\begin{equation}
\lambda_{1,2}={-M \pm \sqrt{M^2+4K^2(1-r^2)} \over 2K(1+r) } 
\label{eq:lambda}
\end{equation}
For small $M$ and $r$ near one, both roots are less than unity in
magnitude, required for the end state to be normalizable.  For the
opposite end of the chain, we use $\psi=\pmatrix{i\cr 1 \cr}$, which
gives solutions which are the inverse of those in Eq.~{\ref{eq:lambda}.

When $r=1$, one eigenvalue vanishes, giving the simple exponential
form in Eq.~\ref{exponential}.  When $M=0$ as well, both eigenvalues
vanish, and the state adheres to a single end site.  If we change the
magnetic field so $\theta\ne \pi/2$, then the energy needs to be
adjusted to keep the two eigenvalues with the same eigenvector.  This
is automatic at zero energy when $\theta=\pi/2$.  The end states cease
to exist when one of the eigenvalues becomes unity in magnitude,
i.e. when
\begin{equation}
M=2Kr.
\end{equation}
For larger $M$ the zero modes are absorbed into the continuum bands.

I now turn to some interesting effects when several of these chains
are connected at a ``junction.''  First note that this simple cross
linked ladder system can have excitations which only move one
direction, backing themselves up into end states.  Start in a state
with all negative energy levels filled and all others empty.  This is
just shy of half filled because the end states are empty.  Now insert
an electron wave packet with wave function $\psi_n\sim\pmatrix {1 \cr
i \cr}$ by applying a superposition of the operators
$\pmatrix{a_n^\dagger & b_n^\dagger\cr} \pmatrix {1 \cr i \cr}$ to the
nearly half filled system.

Because of the phase cancellations, this wave packet initially spreads
only to the left.  It is a mixture of the left end state and positive
energy states, which are effectively plane waves.  As constructed, the
state has no overlap with the right end state.  If we add a damping to
the positive energy modes, the electron will ultimately settle into
the left edge state.  

Now join three such ladders.  Fig.~\ref{fig:icetray5} sketches a
connection of the earlier ladder onto the side of another.  I refer to
the horizontal chain in this figure as the side chain, and the one in
the indicated $z$ direction as the main chain.  Now release a left
moving electron on the side chain as discussed above.  As the state
approaches the junction, it will turn in the direction dictated by the
phases on the main chain.  Asymptotically the electron moves to only
one end of the main chain.

At this point I can make a rather interesting device by generalizing
the model to two component spinors on each atom.  Put a Pauli matrix,
say $\sigma_3$, onto the horizontal bonds on the main chain.  I put no
extra factors on the side chain.  When the particle meets the junction
from the side chain, spin determines the direction it turns.  I have a
spin separator.  If the chain lies in the same physical direction as
the Pauli matrix, it is a helicity filter.  This situation is sketched
in Fig.~\ref{fig:icetray5}.  This is the link to chiral symmetry.

%\section{Domain wall Fermions}

The above picture is at the heart of the domain wall formulation of
chiral fermions.  To map this discussion onto the latter, one ladder
molecule is placed at every site of a three dimensional space lattice.
The system is then half filled, setting the Fermi surface at the level
of the zero modes.  Next, to allow motion in the physical directions,
additional hopping terms couple different sites.  For the standard
picture, corresponding atoms on neighboring space sites are coupled.

To connect with the usual language, the distance along the ladders
represents a ``fifth'' dimension.  The Dirac matrices of the four
dimensional theory become the Pauli matrices inserted in the
``device'' discussed above.  The spin separation of the surface modes
corresponds to their chiral nature.  Since this is a Hamiltonian
discussion, I treat time as a continuous variable \cite{mcih}.
Standard techniques relate this model to the Lagrangian path integral
\cite{transfer}.

On this higher dimensional system, the end states become surface
modes, representing the light Fermions of the physical world.  As in
Fig.~\ref{fig:icetray5}, on each surface these modes have their
helicity projected.  The separation of spin modes corresponds to
chiral symmetry.  The particle energies also receive a contribution
from momentum in the physical directions.  The symmetries force zero
momentum states to zero energy; we have naturally massless particles.
Interactions which renormalize the relative bond strengths will not
change this; any mass renormalization must be proportional to explicit
mass terms.

My phase conventions correspond to a particular representation for the
Dirac gamma matrices.  In conventional lattice gauge language, the
hopping in the extra dimension is a term in the action
proportional to
$$
\overline\psi_n (r+\gamma_5) \psi_{n+1} 
+\overline\psi_{n+1} (r-\gamma_5) \psi_{n}
$$
where $\overline \psi = \psi^\dagger \gamma_0$, the horizontal bonds
in the original model correspond to $\gamma_0\gamma_5$ and the
diagonals to $\gamma_0$.  This plus the corresponding mappings in the
spatial directions gives the Hermitian gamma matrices
\begin{eqnarray}
&&\gamma_0=\pmatrix{0 & 1 \cr 1 & 0\cr}\cr
&&\gamma_5=\pmatrix{0 & -i \cr i & 0\cr}\cr
&&\gamma_i=\pmatrix{0 & -i\sigma_i \cr i\sigma_i & 0\cr}
\end{eqnarray}
The surface states are eigenvalues of $\gamma_5$, again showing the
connection with chiral symmetry.

The magnetic field generating the end states is not in any
``physical'' direction.  It might be thought of as the ``5,6''
component of a field strength tensor $F_{\mu\nu}$, where the fifth
dimension is the extra one of the formulation and the sixth is the
direction of the spinor index.  When the Pauli matrices in the spatial
directions are considered, this becomes a non-Abelian SU(2) field in
the ``$i,6$'' direction.

The crucial point is how, when the extra dimension is large,
symmetries force the fermion mass to zero.  The mass is robust under
renormalizations of the parameters, a property ascribed in continuum
formulations to chiral symmetry.  Indeed, chirality becomes natural on
the lattice, at the expense of an extra dimension.

\acknowledgments
This manuscript has been authored under contract number
DE-AC02-98CH10886 with the U.S.~Department of Energy.  Accordingly,
the U.S. Government retains a non-exclusive, royalty-free license to
publish or reproduce the published form of this contribution, or allow
others to do so, for U.S.~Government purposes.

\input epsf

\begin{figure}
\epsfxsize .7\hsize
\centerline{\epsfbox{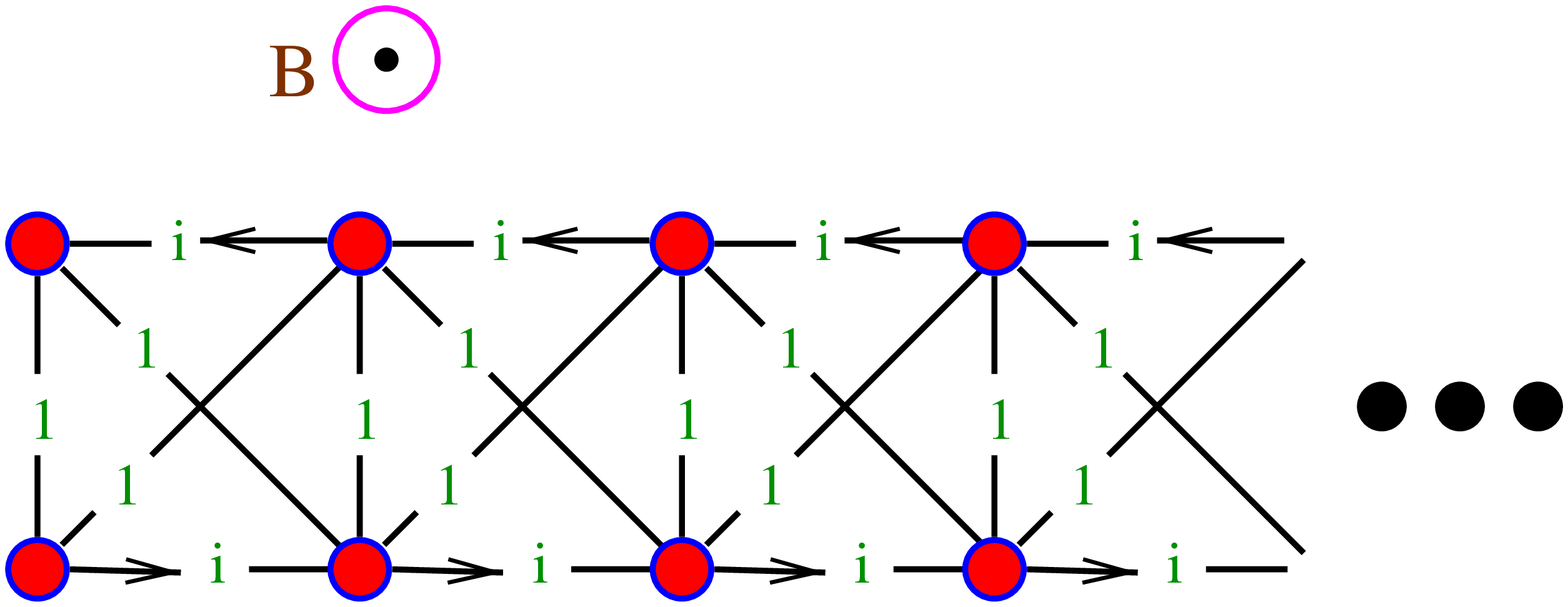}}
\medskip
\caption{The basic cross linked lattice in a magnetic field.  The
numbers on the bonds represent phases giving half a unit of flux per
plaquette.  If we slightly slope the vertical bonds alternately in and
out of the plane, the model is a chain of tetrahedra, linked on
opposite edges.}
\label{fig:icetray2}
\end{figure}

\begin{figure}
\epsfxsize .4\hsize
\centerline{\epsfbox{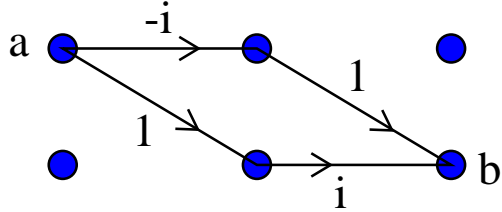}}
\medskip
\caption{With half a unit of magnetic flux per plaquette, the paths
for an electron to move two sites interfere destructively.  
A particle on site $a$ cannot reach $b$.}
\label{fig:cancel}
\end{figure}

\begin{figure}
\epsfxsize .8\hsize
\centerline{\epsfbox{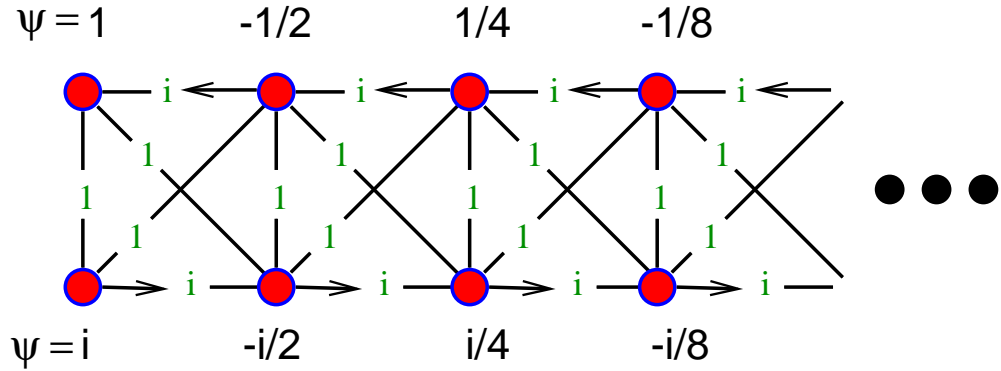}}
\medskip
\caption{A zero energy state bound to the lattice end.}
\label{fig:icetray3}
\end{figure}

\begin{figure}
\epsfxsize .9\hsize
\centerline{\epsfbox{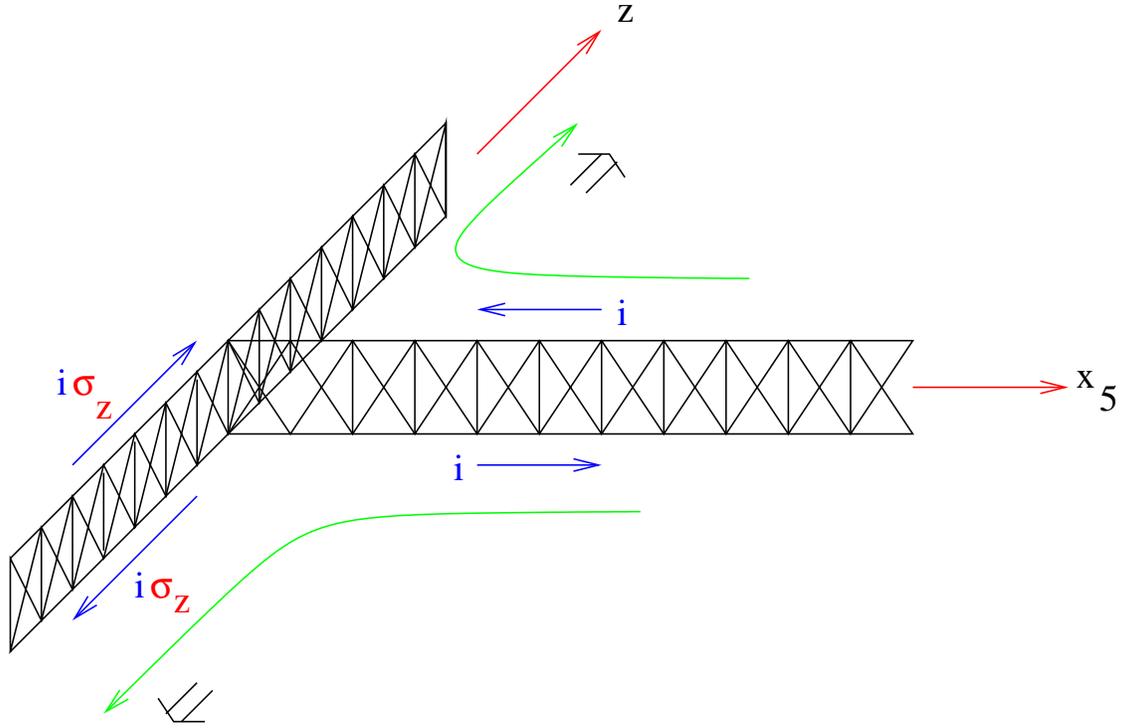}}
\medskip
\caption{Joining a ladder onto the side of another.  When the hopping
on the main chain includes a spin matrix, this is a helicity filter.
A particle traveling down the side chain turns towards the direction
of its spin.}
\label{fig:icetray5}
\end{figure}

\end{document}